\documentclass[format=acmsmall, review=false, screen=true, usenames, dvipsnames, noend, bookmarks, colorlinks, citecolor=green, urlcolor=blue, filecolor=blue, linkcolor=blue]{acmart}

\usepackage{amsmath}

\usepackage[mathletters]{ucs}
\usepackage[utf8]{inputenc}
\usepackage{wasysym}	

\usepackage{hyperref}
\usepackage{algorithmic} 
\usepackage[nameinlink]{cleveref}
\usepackage{natbib}

\usepackage{siunitx}   
\usepackage{makecell}  

\usepackage{amsthm}

\newcommand{\etal}{\hbox{\emph{et al.}}\xspace}

\newcommand{\ie}{\hbox{\emph{i.e.}}\xspace}

\newenvironment{proof-idea}{\noindent{\bf Proof Idea}\hspace*{1em}}{\bigskip}

\theoremstyle{plain}

\theoremstyle{remark}

\theoremstyle{definition}

\DeclareUnicodeCharacter{183}{\cdot}						
\DeclareUnicodeCharacter{931}{\ensuremath\Sigma}			
\DeclareUnicodeCharacter{9001}{\ensuremath\langle}			
\DeclareUnicodeCharacter{9002}{\ensuremath\rangle}			
\DeclareUnicodeCharacter{9608}{\ensuremath\blacksquare}		
\DeclareUnicodeCharacter{1013}{\in}							
\DeclareUnicodeCharacter{8213}{---}							

\usepackage{listings}
\usepackage{color}
\usepackage{tikz}
\usepackage{amsthm}
\usepackage{amssymb}
\usepackage{stmaryrd}
\usepackage{graphicx}
\usepackage{mathpartir}
\usetikzlibrary{arrows,shapes,snakes,automata,backgrounds,petri,shadows}
\usepackage{enumitem}
\usepackage{pifont}
\usepackage[vlined,ruled,linesnumbered]{algorithm2e}
\usepackage{float}
\usepackage{framed}
\usepackage{siunitx}
\usepackage{xspace}
\usepackage{microtype}

\definecolor{codegreen}{rgb}{0,0.6,0}
\definecolor{codegray}{rgb}{0.5,0.5,0.5}
\definecolor{codepurple}{rgb}{0.58,0,0.82}
\definecolor{backcolour}{rgb}{0.95,0.95,0.92}
\definecolor{pblue}{rgb}{0.13,0.13,1}
\definecolor{pgreen}{rgb}{0,0.5,0}
\definecolor{pred}{rgb}{0.9,0,0}
\definecolor{pgrey}{rgb}{0.46,0.45,0.48} 

 \lstdefinestyle{mystyle2}
 {
  	showspaces=false,
  	showtabs=false,
  	breaklines=true,	
  	showstringspaces=false,
  	breakatwhitespace=true,
  	commentstyle=\color{pgreen},
  	keywordstyle=\color{pblue},
  	stringstyle=\color{pred},
  	basicstyle=\ttfamily,
  	moredelim=[il][\textcolor{pgrey}]{},
  	moredelim=[is][\textcolor{pgrey}]{\%\%}{\%\%}
}

\lstdefinestyle{mystyle}
{
    commentstyle=\color{codegreen},
    keywordstyle=\color{magenta}\bfseries,
    numberstyle=\tiny\color{codegray},
    stringstyle=\color{codepurple}
}

\newtheorem{definition}{Definition}[section]

\theoremstyle{remark}

\newcommand\sem[1]{\ensuremath{\llbracket #1\rrbracket}}
\newcommand\semc[1]{\ensuremath{\llbracket #1\rrbracket}^c}
\newcommand{\false}{\mathsf{false}}
\newcommand{\true}{\mathsf{true}}

\newcommand{\mysucc}{\mathsf{succ}}

\newcommand{\stmt}{s}

\newcommand{\divergesymb}{\bot}

\newcommand{\bugDens}{\mathtt{BD}}

\newcommand{\ourAnalysis}{Cook\xspace} 
\newcommand{\ourTool}{Endeavour\xspace}
\newcommand{\rep}{\mathcal{R}}

\newcommand{\hslashslash}{%
  \raisebox{.9ex}{%
    \scalebox{.7}{%
      \rotatebox[origin=c]{18}{$-$}%
    }%
  }%
}
\newcommand{\dslash}{%
  {%
   \vphantom{d}%
   \ooalign{\kern.05em\smash{\hslashslash}\hidewidth\cr$d$\cr}%
   \kern.05em
  }%
}

\newcommand{\ns}[1]{\todo{Nassim: #1}}

\begin{document}

\title {Sub-Turing Islands in the Wild}

\author{Earl T. Barr}
\affiliation{\institution{University College London} \city{London} \country{UK}}
\author{David W. Binkley}
\affiliation{\institution{Loyola University Maryland} \city{Baltimore} \country{USA}}
\author{Mark Harman}
\affiliation{\institution{Facebook and University College London} \city{London} \country{UK}}
\author{Mohamed Nassim Seghir}
\affiliation{\institution{University College London} \city{London} \country{UK}}





\begin{abstract}

Recently, there has been growing debate as to whether or not static analysis
  can be truly sound.  In spite of this concern, research on techniques seeking
  to at least partially answer undecidable questions has a long history.
  However, little attention has been given to the more empirical question of
  how often an exact solution might be given to a question despite the question
  being, at least in theory, undecidable. This paper investigates this issue by
  exploring \emph{sub-Turing islands} --- regions of code for which a question
  of interest is decidable.  We define such islands and then consider how to
  identify them. We implemented \ourTool, a prototype for finding
  sub-Turing islands and applied it to a corpus of 1100 Android applications,
  containing over 2 million methods. Results reveal that 55\% of the all
  methods are sub-Turing. Our results also provide empirical, scientific evidence
  for the scalability of sub-Turing island identification. 

Sub-Turing identification has many downstream applications, because islands
  are so amenable to static analysis. We illustrate two downstream uses of the
  analysis. In the first, we found that over 37\% of the verification
  conditions associated with runtime exceptions fell within sub-Turing islands
  and thus are statically decidable.  A second use of our analysis is during
  code review where it provides guidance to developers. The sub-Turing islands
  from our study turns out to contain significantly fewer bugs than ``the
  swamp'' (non sub-Turing methods). 
The greater bug density in the swamp is unsurprising;  the fact that bugs remain
prevalent in islands is, however, surprising:  these are bugs whose
 repair can be fully automated.

\end{abstract}



\maketitle








\section{Introduction}
\label{sec:intro}
This paper seeks answer the following fundamental question at the intersection of programming languages theory and empirical software engineering:
\begin{quote}
What portion of the code of a large corpus of real software systems lies in Sub-Turing islands; `islands' of code that denote computation for which interesting program analysis questions are decidable?
\end{quote}
We use the term `Turing Swamp' to refer to any code that does not lie in such  a
Sub-Turing island. Of course, merely determining whether or not code lies within
an island or in the swamp is, itself, undecidable. 
Therefore, our tool uses a simple conservative under-approximation of
Sub-Turing islands (and corresponding over-approximation of the swamp).

Our Sub-Turing island identification algorithm, Cook, guarantees that the
halting problem is decidable for any computation it identifies as lying within
an island; as a result, Cook necessarily under-approximates the amount of
code that lies within such islands.
Even with this relatively simple under-approximation,  we were able to determine that a large proportion of non-trivial production code (for Android) does indeed lie in island (not swamp) code. That is, for a corpus of 1100 Android applications, containing over 2 million methods, we found that 55\% of the methods are sub-Turing. 

Even if we remove the `long tail' of simple methods (like getters and setters and methods with fewer than 30 bytecode instructions) we still find that 22\% of all code lies in a Sub-Turing island. We then ask
\begin{quote}
Since we find that at least one fifth of non-trivial real-world systems lies in a Sub-Turing island, what are some of the ramifications for programming languages and software engineering applications that rely on static analysis?
\end{quote}

To investigate these implications we conducted two empirical studies of the impact for sub-Turing islands. Even with our conservative under-approximation, we found that (at least) 37\% of the verification conditions for runtime exceptions (e.g., array  bounds and null pointer violations) lie within sub-Turing islands. Furthermore, (for a dataset of ten open source applications), we found a statistically significant difference in bug density, with a large effect size. 

These findings reveal a glimpse of the potential implications and applications of Sub-Turing analysis.
In a single paper we cannot claim to have addressed more that the first few natural questions that occur when considering the approximate computation of the boundary between Sub-Turing islands and the swamp.
Nevertheless, we believe that our results demonstrate that a surprisingly large portion of code does clearly lie within a Sub-Turing island and that there is practical merit in studying islands to inform and improve static analysis.
Sub-Turing Islands support fully automatic and precise symbolic reasoning; this
reasoning might be exploited for bug repair and free humans to concentrate
problems occurring in the swamp.

There are many avenues for future work. We outline some of these and their relationship to existing trends of intellectual investigation in the programming languages and software engineering research communities.
We hope that this paper will stimulate the further investigation of Sub-Turing analyses of software and real-world applications of these findings.
Our paper seeks to motivate this research agenda with scientific evidence for the prevalence of Sub-Turing islands (within Android applications in this case) and the real-world impact and implications for bug density and verification. 

Specifically, the contributions of this paper are the following.
\begin{itemize}
\item We introduce and formalise the concept of sub-Turing island.
\item We provide an analysis for identifying sub-Turing islands and its implementation in the prototype tool \ourTool.
\item We reveal that Sub-Turing code is more prevalent in real-world system than might be expected: a conservative lower bound is at least one fifth of non-trivial Android App code is Sub-Turing.
\item We demonstrate that Sub-Turing island analysis has great potential for real-world application. Specifically, we report that 37\% of the array bounds and null pointer verification conditions lie within islands, while islands enjoy 
lower bug density than the Turing swamp.
\end{itemize} 
 
\section{Sub-Turing Islands} 
\label{sec:st-islands}

This section first defines \emph{Sub-Turing Island} where the definition is
parameterized by a decision procedure.
As an illustrative decision procedure, we consider terminating islands,
Sub-Turing islands with a conservative decision procedure for halts versus may not halt. 
The bulk of this section then presents the syntax and semantics of Carib, a
core language that facilitates the identification of Sub-Turing islands.

\begin{definition}[Sub-Turing Island]
\label{def:stisland1}

A region of code $r$ is \emph{sub-Turing} with respect to property $p$ if there
exists a decision procedure $\mathsf{D}(p,r)$ that determines whether $p$ holds
over all executions of $r$. 

\end{definition}

Under Rice's theorem~\cite{rice53}, Sub-Turing islands are only computable when
$\mathsf{D}$ approximates $\neg p$ while being certain when it determines $p$ holds;  
it cannot decide both $p$ and $\neg p$. For an arbitrary property, a suitable
decision procedure may not exist, hence the existential quantification in the
definition of sub-Turing.  
Finding a decision procedure is based on human ingenuity.  
The parameterisation of the definition on decision procedure $\mathsf{D}$, implies
that sub-Turing islands are only defined with respect to a given decidable property.  
Different static analyses can safely approximate the islands, and with
different levels of precision, thereby giving rise to the generation of different islands.
However, if any approximation safely under-approximates the code that lies
within an island  then it will be safe to make `island-aware assertions and
inferences' within any given island. 

We focus our investigation on a decision procedure for the halting problem.  
Our realisation, given in \autoref{sec:loops}, soundly approximates this undecidable problem.
Given this decision procedure, we frame the \emph{Terminating Islands Identification Problem}
in terms of states, $\sigma : L \rightarrow V_\bot$, where $L$ denotes the set of
program l-values and $V_\bot$ denotes the lifted value domain $V$, which we
leave otherwise unspecified.
We say that a state, $\sigma$ is \emph{divergence free} if none of its l-values
are mapped to $\bot$.  
Given a divergence-free state $\sigma$ and a region of code $r$, our goal is to
determine if $\semc{r} \sigma$ is also divergence-free; we formalize $\semc{r}
\sigma$, the semantics of Carib, in \autoref{sec:carib}.  
From a practical perspective, we are interested in regions that represent
meaningful code fragments such as a method or a loop body.  

\begin{definition}[The Terminating Island Identification Problem]
\label{def:identify_STI}

Given a region of code $r$, the \emph{Terminating Island Identification Problem}
is to determine if 
\[\forall \sigma \in \Sigma.\; (\forall x \in L.\;\sigma[x] \neq \bot
\;\Rightarrow\; \forall x \in L.\; (\semc{r} \; \sigma ) [x] \neq \bot).
\]
\end{definition}
\noindent When this condition is satisfied, divergence can only result from the
code that makes up $r$.  In this sense, $r$ is independent of its context (the
rest of the program).  Since we consider only a decision procedure for
termination in this paper, we use Sub-Turing Island to refer to a Terminating
Island in what follows.

To illustrate the goal of our analysis, consider the three examples shown in
\autoref{fig:examples2} where the region of code considered is a method.  To
being with, method \lstinline{foo} of \autoref{fig:examples2}a calls method
\lstinline{bar}, which is clearly sub-Turing; thus it is not a source of
divergence to its caller \lstinline{foo}, which is also sub-Turing.  In
contrast, in \autoref{fig:examples2}b, callee \lstinline{bar} is not sub-Turing
as it contains a loop whose termination can not be guaranteed.  As a result its
caller \lstinline{foo} is also not sub-Turing.  Finally,
\autoref{fig:examples2}c is similar to \autoref{fig:examples2}b except that the
source of divergence is a call to an API method, which may diverge (A call
to a recursive method would have the same effect.)  The potentially divergent
API call does not, however, relegate Line 4 to the swamp, despite its control dependence
on the call to \lstinline+bar+.  This is because a Carib function always
terminates, 
so the fact that a Line 4 is termination-sensitive control dependent
on Line 3 does not matter \cite{sdetal:unifying, kaetal:acm-surveys}.

\begin{figure*}[t!]
\begin{center}
\begin{tabular}{c@{\hspace{0.5in}}c@{\hspace{0.5in}}c}
\hline
\\
  \begin{lstlisting}[escapechar=\%]
void foo(){%{\;\large \color{OliveGreen}\ding{52}}%
  int x = 5; 	      
  x = bar(); 
  return x;	
}  
    
int bar(){%{\;\large \color{OliveGreen}\ding{52}}%
  int y = 1;
  return y;      
}
  \end{lstlisting}

&
  \begin{lstlisting}[escapechar=\%,mathescape=true]
void foo(){%{\;\large \color{red}\ding{56}}%
  int x = 5; 
  x = bar(); 
  return x;      
}  
    
void bar(){%{\;\large \color{red}\ding{56}}%
  int y = 0;
  while($C$){
    y = y+1;
  } 	      
  return y;	
}
  \end{lstlisting}

&
  \begin{lstlisting}[escapechar=\%]
void foo(){%{\;\large \color{OliveGreen}\ding{52}}%
  int x = 5; 
  unused = bar(); 
  return x;      
}  
    
void bar(){%{\;\large \color{red}\ding{56}}%
  int y = 0;
  r = api();
  if (r)
    y = y+1;
  } 	      
  return y;	
}
\end{lstlisting}


\\
(a)  & (b) & (c) 
\\
\hline

\end{tabular}
\end{center}

\caption{Three examples illustrating the outcome of our Sub-Turing analysis. Symbol \ding{52} indicates that the method is sub-Turing and \ding{56} indicates the opposite. 
In the example $C$ is a condition that cannot be proven to eventually be false
while \lstinline+api+ is a call to an unknown API method.}
\label{fig:examples2}
\end{figure*}

\subsection{Carib:  Its Syntax and Semantics}
\label{sec:carib}

\begin{figure*}[t]
\begin{center}
\[
\begin{array}{rcl}
%
\textsc{prog} &::= &  \textsc{method},\cdots,\textsc{method} \\
\textsc{method} &::= & id(id,\cdots,id)\;\textsc{stmt} \\   
\textsc{stmt} &::= &  \textsc{stmt};\textsc{stmt} \;|\; \textsc{assign} \;|\; \textsc{ife} \;|\; \textsc{while} \;|\; \textsc{call} \;|\; \mathsf{return}\;  id\\
\textsc{assign} &::= & id := c \;|\; id :=  id \;|\; id := \mathsf{op}\;id \mid
	id := id \;\mathsf{op}\;id \\
	    & & id := id.id \mid id.id := id \mid id :=  id[id] \;|\; id[id] :=  id \\
\textsc{ife} &::= &  \mathsf{if}\;\textsc{cond}\;\mathsf{then}\;\textsc{stmt}\;\mathsf{else}\;\textsc{stmt} \\
\textsc{while} &::= &  \mathsf{while}\;\textsc{cond}\;\mathsf{do}\;\textsc{stmt}\\
\textsc{call} &::= &  id := id(id,\cdots,id)\\
\textsc{cond} &::= & id \;\mathsf{rel\_op}\; id 
\end{array}
\]
\end{center}
  \caption{The grammar $\mathcal{G}$ for Carib, our core Jimple-like language.}
\label{fig:carib}
\end{figure*}

\autoref{fig:carib} defines $\mathcal{G}$, the grammar of Carib, our core
language.  Carib is a Jimple-like~\cite{Vallee-RaiGHLPS00} intermediary
representation with a minimal set of instructions. Vall\'ee \etal
\cite{Vallee-RaiGHLPS00} and Bartel \etal \cite{BartelKTM12} have shown that
Jimple can encode the entire instruction set of widely deployed virtual
machines, such as the JVM and Dalvik.  A Carib program is a set of methods
derived from the nonterminal \textsc{prog}.  Carib incorporates three
simplifications that ease the presentation.  First, instead of the
usual syntax for method invocation $o.m(\cdots)$, Carib uses the form
$m(o,\cdots)$, with the receiver being the first argument.  Second, Carib defines
only the two structured control constructs \textsf{while} and \textsf{if-else}.
Finally, to simplify reasoning about side-effects, Carib restricts pointer
dereferences to its assignment statement, where the dereference operator can
only appear either in the LHS or RHS \emph{alone}.  Further, the call syntax
only permits an $id$ as an actual parameter, again ruling out pointer
dereferences.  These properties simplify reasoning about aliasing in Carib.

Carib's semantics, $\semc{\cdot}$, extends a conventional semantics,
$\sem{\cdot}$, such as Winskel's IMP~\cite{Winskel} where $\sem{s}$ is a
partial function from $\Sigma$ to $\Sigma$ that updates the state to reflect
the execution of $s$.  
For $s \in L(\mathcal{G})$ (\autoref{fig:carib}), $\semc{s}$
is identical to the conventional semantics, $\sem{s}$, when $s$ terminates.
When $s$ does not terminate, $\semc{s}$ reifies the nontermination by binding
$\bot$, to each variable modified by $s$.  

To reify nontermination, $\semc{s}$ must first identify it.
In Carib there are three potential sources: loops, recursive method calls, and
(unknown) API calls, 
The semantics uses three oracles in the identification: $O_t$, $w$, and $A$
(\autoref{sec:realising_oracles} discusses our computable approximations to
these three).
For example, the termination oracle, $O_t$ is used to identify nonterminating
loops as well as recursive methods that may diverge.
\autoref{fig:carib:params} defines these oracles and other symbols and
functions used to define Carib's semantics.

Finally, we formalise the notions of state and state update as used in the
Carib semantics. 
As a convenience, we assign a unique name to each local variable and
formal parameter, and then simply refer to only those names that are in scope.  
We use $L$ to denote the set of all program l-values (in \autoref{fig:carib}, 
these include identifiers $id$ and array/structure references, $id[id]$).
An l-value denotes a memory location that holds a value from the lifted value
domain, $V_\bot$, where $\bot$ denotes divergence; we leave $V_\bot$ otherwise
unspecified.
A program state $\sigma: L \rightarrow V_\bot$ maps each l-value
$x$ to its value $v_x$.  
$\Sigma$ denotes the (possibly infinite) set of all program states.  
We write $\sigma[x]$ to denote the value that $\sigma$ maps $x$ to and
$\sigma[x:=v]$ to denote the updated state $\sigma'$ where $\sigma'[x] = v$
and $\sigma'[y] = \sigma[y], x \ne y$.
As a notational convenience, we write $\sigma[X]$ for variable set $X$ to denote 
$\{\sigma[x] \;|\; x \in X\}$ and $\sigma[\overrightarrow{Y}:=\overrightarrow{Z}]$ as
shorthand for $\sigma[y_1:=\sigma[z_1],\cdots, y_k:=\sigma[z_k]\;]$, with
$\overrightarrow{Y} = (y_1,\cdots,y_k)$ and $\overrightarrow{Z} =
(z_1,\cdots,z_k)$ and $y_i,z_i \in L$.  
Finally, we write $\sigma[\overrightarrow{Y} := v]$ to denote
$\sigma[y_1:=v,\cdots,y_k:=v]$.  

Carib's semantics must account for two potentially divergent constructs,
\textsc{while} and \textsc{call}.  
If a loop does not terminate, our semantics effectively replaces the loop
with a parallel assignment of $\bot$ to all the l-values that
the loop may modify.  
We handle recursive calls in the same way.  
Other constructs in $L(\mathcal{G})$ may propagate $\bot$, but will not
introduce it.  
In essence, our semantics is a collecting semantics based on taint
analysis where \textsc{while} and \textsc{call} are the only taint sources. 
We say that a program point is in the \emph{swamp} if $\bot$ reaches it,
otherwise the point is a sub-Turing island. 
Finally, we emphasis that, under its semantics Carib methods always terminate. 

\begin{figure}[t]
\begin{center}
\begin{tabular}{ll}
  $w(s)$ & set of all l-values potentially written by the execution of $s$ \\
  $O_t(s)$  & the termination oracle used to check for nontermination of $s$ \\
  $A$    & the set of (assumed divergent) API methods \\
  $R$    & the set of recursive methods $O_t$ deems divergent \\
  $\mathit{fv(e)}$ & the set of free variables in expression $e$  \\
  $\mathit{ret}$  & a fresh pseudo-variable used to hold a method's return value
\end{tabular}
\end{center}
\caption{Symbols and functions used to define Carib's semantics in
	\autoref{fig:rules}; the first three are parameters.}
\label{fig:carib:params}
\end{figure}

\begin{figure}[t]
  \begin{center}
    
\setlength\tabcolsep{2.5pt}
\begin{tabular}{rcl}

	\textsc{sequence} : $\semc{s_1;s_2}$ & $=$ & $\lambda\sigma.\;  \semc{s_2}(\semc{s_1} \; \sigma ) $

\\
\\
\textsc{assign} : $\semc{s = v := e}$ & $=$ & $\begin{cases}
        \lambda\sigma.\; \sigma[\overrightarrow{w(s)}:=\bot] & \text{if } \bot \in \sigma[\mathit{fv}(e)]  \\   
	\sem{s} & \text{otherwise} \\
 \end{cases} $

\\
\\
\textsc{ife} : $\semc{s =
\mathsf{if}\;e\;\mathsf{then}\;s_1\;\mathsf{else}\; s_2}$ & = & $\begin{cases}
        \lambda\sigma.\; \sigma[\overrightarrow{w(s)}:=\bot]  & \text{if } \bot \in \sigma[\mathit{fv}(e)]  \\   
	\lambda\sigma.\; \sem{e} \; \sigma \;?\; \semc{s_1} \; \sigma  : \semc{s_2} \; \sigma  & \text{otherwise} \\
 \end{cases} $

\\
\\
\textsc{while} : $\semc{s = \textsc{while}\;e\;\mathsf{do}\; s_1}$ & $=$ & $\begin{cases}
        \lambda\sigma.\; \sigma[\overrightarrow{w(s)}:=\bot]  & \neg O_t(s)  \vee \bot \in \sigma[\mathit{fv}(e)]  \\   
	\lambda\sigma.\;  \sem{e} \; \sigma \;?\; \semc{s_1
	;\textsc{while}\;e\;\mathsf{do}\; s_1} \; \sigma  : \sigma& \text{otherwise} \\
 \end{cases} $

\\
\\
\textsc{call} : $\semc{s = r := m(\overrightarrow{X})}$ & $=$ &
$\begin{cases}
    \lambda\sigma.\; \sigma[\overrightarrow{w(s)}:=\bot, r := \bot]  & m \in A \cup R  \\   
  \lambda\sigma.\;\mathrm {let ~} m \textrm{~be~} m(\overrightarrow{Y}), \; \sigma' = \semc{\mathit{body}(m)} \; \sigma[ \overrightarrow{Y} := \overrightarrow{X}]  \\
   \mathrm{ \quad\>\>\>~~in~~ } \sigma[r := \sigma'[ret] \; ] & \text{otherwise} \\
  \textrm{where~} \mathit{body}(m) \textrm{~denotes the body of method,~} m.  
 \end{cases} $
\\

\end{tabular}
\end{center}

\caption{Carib's denotational semantics: \textsc{while} and
\textsc{call} can introduce divergence ($\bot$); the other equations
only propagate it.}
\label{fig:rules} 
\end{figure}

\autoref{fig:rules} presents Carib's semantics, $\semc{.}$.  The rule for
\textsc{while} leverages $O_t$ to determine if a loop terminates.  For a loop
$s$ that may not terminate, the first line of \textsc{while} rule binds $\bot$
to each l-value potentially written during the execution of $s$.  The externally
supplied function $w(s)$ identifies these l-values.
\autoref{sec:realising_oracles} describes our conservative realisation of $O_t$
as well as our conservative determination of the set of written l-values.

The second source of non-termination is calls to recursive methods and API
methods, denoted $R$ and $A$ in \autoref{fig:rules}.  For the \textsc{call}
rule, the first case binds $\bot$ to all l-values that the called
method's execution potentially updates together with $r$, the variable receiving
the method's return value.  In \textsc{call}'s second case, conventional
semantics apply.  
Here, $\mathit{body}(m)$ denotes the statements of the called method.  
Working outward from $\semc{\mathit{body}}$, we evaluate $m$'s
$\mathit{body}$ on the state formed by binding $m$'s formals to the actuals
found in the call.  Other than the return value, there is no need to map
information back to the caller because $L(\mathcal{G})$ uses call by value
semantics.  In the case of objects (and arrays), Carib passes a copy of the
reference to the object to the callee thereby allowing the called method to
update (only) the members of the class (or array) associated with the actual.  
To handle return values, we store the value in $\sigma'[\mathit{ret}]$ and then
update the final state to bind $r$ to this value.

Finally, the \textsc{ife} and \textsc{assign} rules can only propagate $\bot$;
they do not introduce it.  
For \textsc{assign}, if $\bot$ reaches any variables in $e$, \textsc{assign} binds $\bot$ to $v$; 
otherwise, the conventional semantics $\sem{s}$ is used. 
When $\bot$ reaches an \textsf{if} statement's conditional expression the \textsc{ife} rule 
assigns $\bot$ to all l-values potentially written by either branch of the if statement;
otherwise, it applies $\semc{\cdot}$ to the appropriate branch.

\section{Realising Oracles and Translating to Carib}
\label{sec:realising_oracles}

To analyse industrial programs, we must translate them into Carib and
instantiate Carib's three oracles: its writable location oracle $w$, its
termination oracle $O_t$, and its externally defined, set of divergent API
methods $A$.  To handle constructs Carib does not define, we translate to Carib in the
obvious way, conceptually de-sugaring them.  For $A$, we assume a user-supplied
list.  Below, we describe how we realize $w$ and $O_t$ for Android bytecode.
Realising these oracles is not enough.  To apply the Cook analysis to actual
programs, we also need to change the semantics of their potentially divergent
constructs, like loops and method calls.  We achieve this via a program
transformation $\phi$ that replaces each potentially divergent construct $c$
with parallel assignments of $\bot$ to the l-values of $w(c)$.

\subsection{Soundly Identifying Potential Writes}
\label{sec:aliases}

Realising $w$ requires finding writable locations, both syntactic l-values and
what can be reached through them.  We currently harvest syntactic l-values
from assignment statements without considering their feasibility.  
Computing reachable l-values would require handling aliasing, which occurs when
two l-values refer to the same object.  
To account for the possibility of aliasing in the analysis we use l-value
\emph{representatives} where two l-values are aliases if they have the same
representative.  

We want our \ourAnalysis analysis (\autoref{sec:discovery}) to scale to large
applications, so we need a sound and efficient handling of aliases. Instead of
performing pointer analysis (\autoref{sec:pointer_analysis}), we use Sundaresan
\etal's approach~\cite{SundaresanHRVLGG00} as it offers a simple and scalable
solution.  Their approach is based on the observation that only objects of
compatible types can be aliases in a type-safe programming language, such as
Java.  Another advantage of this approach is its simplicity and ease of
implementation.  Sundaresan \etal use a flow-insensitive analysis because their main
purpose is to keep track of object types. In our case, we need to track value
transfer between variables. Therefore, we take inter-variable flow relations
into account. For example, in the code \lstinline{x := y; y := z}, a
flow-insensitive analysis captures data-flow from \lstinline{y} to
\lstinline{x} and \lstinline{z} to \lstinline{y} and the spurious
flow from \lstinline{z} to \lstinline{x}.  Our approach does not include the
spurious flow. Despite this additional precision, we still benefit, in terms of
scalability, from Sundaresan \etal's alias handling.  

Let $L_T$ denote $L$ augmented with abstracted locations created from the types
of the subject program. 
Formally, we define $\rep: L \rightarrow L_T$ to map each l-value to its \emph{representative}:
\begin{equation*}
	\rep(l) = 
	\begin{cases}
		T(o,f).f 	& \text{for field reference } l := o.f \\
		\rep(a) 	& \text{for array reference } l := a[i] \\
		l 		    & \text{otherwise (\ie, scalar references)}\\
	\end{cases}
\end{equation*}
where $o$ is an object of type $t$, $f$ is a field, $a$ is an array,
$i$ is an index, and $T(o,f)$ denotes the highest class in the type hierarchy of
$t$ that contains the field $f$.  

In Carib (\autoref{sec:st-islands}), an
l-value is either a variable, an array, or a field access.  In the absence of an
array or field dereference, the mapping is simply $\rep(x) = x$.  The other two
cases, a field or array access, are more involved.
Two field accesses $o_1[f]$ and $o_2[f]$ are aliases if $o_1$ and $o_2$ point to
the same object. To handle this case, all potentially aliasing field accesses
must have the same representative. In type safe languages, like Java, $o_1[f]$
and $o_2[f]$ can be aliases only if $o_1$ and $o_2$ belong to the same type
hierarchy.   
While in principle if $o_1[f]$ and $o_2[f]$ are aliases, then either suffices
as the representative, for ease of identification, we include in $\rep$'s range 
representatives based on type names, specifically $T(o, f)$.  

An array access l-value can alias for two reasons: reference and index.  In the
reference case, $a[i]$ and $b[i]$ alias if $a$ and $b$ alias.  Alternatively,
$a[i]$ and $a[j]$ alias if $i = j$.  To take both into account, we perform a
lightweight alias analysis that partitions array terms into parts of potential
aliases. Given an array $a$, $A(a)$ returns the representative of $a$'s alias
part. Defining the representative of $a[i]$ as $A(a)[i]$ solves the problem of
reference-induced aliasing, but not index-induced aliasing.  Tracking indexes
may generate an unbounded number of terms when indices are modified inside a
loop. Therefore, $\rep$'s third case conservatively assumes all indices alias.

Using $\rep$ and $\mathit{lvals}(s)$, the set of all syntactic l-values 
in $s$, we approximate $w$ as
\begin{align*}
  \tilde{w}(s) = \bigcup_{x \in \mathit{lvals}(s)} \rep(x).  
\end{align*}	
This realisation of $w$, $\tilde{w}$, assumes $\mathit{lvals}$ can access $s$'s
internals.  This assumption does not hold, in general, for API calls whose
implementation can be externally defined in a black box library.  Such API
calls are prevalent in real world code.  To handle them, we use a second
instantiation of $w$ that we describe in \autoref{sec:rewrite:dconstructs}
where we first use it.

\subsection{Identifying Divergent Loops and Calls}
\label{sec:loops}

Realising Carib's semantics for real bytecode demands that we first identify
loops.  Since bytecode permits unstructured loops, we implemented a loop
detection analysis that searches for loops in the control flow graph of each
method, which was experimentally shown to outperform existing
alternatives~\cite{WeiMZC07}. 
This analysis is an optimised depth first traversal that discovers the control flow graph.  
This analysis also allows us to detect both
simple and complex loops including nested loops and those constructed using
\textsf{goto}s.  A purely syntactic method, this analysis is complete (finds all
loops), but unsound, in that it may report infeasible loops.

Having identified loops, we turn to realizing $O_t$.  Despite the
undecidability of loop termination in general, we can statically determine that
some loop forms terminate. While simple, our oracle realisation is not purely
syntactic. It simplifies a loop before checking its form. Let $\omega$ be a
loop that does not contain any nested loops, but otherwise has an arbitrary
body. The execution of loop $\omega$ can be expressed as a sequence of
single-iteration cycles: $C_1,\ldots,C_k$. Our oracle concludes that $\omega$
terminates iff both of the following hold: 1) each cycle increments a counter
and 2) this counter is bounded in each cycle.  These two conditions guarantee
the existence of an increasing ranking function that is bounded from above,
which is sufficient to ensure loop termination~\cite{PodelskiR04}. This
instantiation of $O_t$, which we denote $\tilde{O}_t$,
safely and conservatively under-approximates the set of
terminating loops.

Our corpus of Android apps includes 627,423 loops.  Of these, our oracle
shows that 330,894 (53\%) terminate.  Despite its simplicity and
conservatism, $\tilde{O}_t$ identifies a large number of
terminating loops.

In \autoref{sec:carib}, we use $O_t$ to populate $R$, the set of divergent
recursive methods.  $\tilde{O}_t$ works only on loops, which necessitates a
separate mechanism for $R$. To do so, we conservatively build a call graph for
the program using the class hierarchy approach~\cite{SundaresanHRVLGG00} that
provides a conservative approximation of the runtime types of receiver objects.
We then identify recursive methods as those nodes belonging to strongly
connected components in the call graph. To this end, we use Tarjan's algorithm
for detecting strongly connected components~\cite{Tarjan72}.

\subsection{Rewriting Divergent Constructs}
\label{sec:rewrite:dconstructs}

\begin{figure}[t]
\begin{align*}
  \textsc{loop}: s = \mathsf{while}\;c\;\mathsf{do}\; s_1 \wedge \neg \tilde{O}_t(s)\; 
	& \Rightarrow \; s \;   \rightarrow \; \rep(\vec{Y}) := \divergesymb 
  \text{ where } \vec{Y} = \tilde{w}(s) \\
\textsc{rec}: m \in R \wedge s =  r := m(\vec{X})\; 
	& \Rightarrow \; s \;\rightarrow \;
  \rep(\vec{Y}) := \divergesymb \text{ where } \vec{Y} = \tilde{w}(\mathit{body}(m)[\vec{X}/\vec{F}]) \\
\textsc{api}: \mathsf{m} \in A \wedge s = r:= \mathsf{m}(\vec{X})\; 
	& \Rightarrow \; s \;\rightarrow \; \rep(\vec{Y}) := \divergesymb;
	\rep(r) := \divergesymb
  \text{ where } \; \vec{Y} = \bigcup_{x_i \in \vec{X}} \mathit{RLV}(x_i)
\end{align*}
	\caption{The program transformations $\phi$ used to convert divergent
  constructs into parallel assignments of $\bot$; $\vec{F}$ used in \textsc{REC} is $m$'s formals.}
	\label{fig:phi}
\end{figure}

To track divergence in a program, we identify divergent constructs and replace
them with assignments of $\divergesymb$ to every l-value representative 
potentially modified. We do this as a source-to-source transformation $\phi:
L(\mathcal{G}) \longrightarrow L(\mathcal{G})$, which maps each statement
$\stmt$ to a statement $\stmt'$ that explicitly includes all necessary
assignments of $\bot$. As \autoref{fig:phi} depicts, we define $\phi$ using the
three rewriting rules: \textsc{loop}, \textsc{rec}, and \textsc{api}.
The \textsc{loop} rule replaces the loop with a parallel assignment of $\bot$
to all l-value representatives that the loop modifies.  
It uses $w$ to identify these l-values.  
The \textsc{rec} rule does the same for calls to recursive methods.  

Handling loops and recursion makes us complete with respect to our language
Carib (\autoref{sec:st-islands}), assuming that a program is self-contained (\ie
does not make API calls).  Most programs, however, make API calls to external
libraries.  Because an external library is a black box, we cannot
syntactically determine which of its parameters it may write through.  Thus, we
need a second instantiation of $w$ to handle API calls.  This second
instantiation determines all the l-value representatives potentially modified
using a given formal parameter.  
For arrays or objects, however, we must consider their fields.  Let
`\textsf{fields}' be the set of fields of a variable where the empty set denotes
a scalar.  Given the formal parameter \lstinline{x}, its reachable l-values are
the set of l-value representatives given by the function $\mathit{RLV}$ defined
as follows
\begin{align*}
\mathit{RLV}(x) = \{\rep(x[f]) \;|\; f \in \mathsf{fields}(x)\} \bigcup_{f \in \mathsf{fields}(x)} \mathit{RLV}(x[f]).  
\end{align*}
$\mathit{RLV}$ is irreflexive because, under Carib's call by value semantics,
actual parameters are immutable.  In \autoref{fig:phi}, the \textsc{api} rule uses
$\mathit{RLV}$ to handle API calls where the set $\vec{Y}$ includes all l-value
representatives reachable from $m$'s \emph{actual} parameters, as determined by
$\mathit{RLV}$.

As an illustration of $\mathit{RLV}$, consider the code shown in
\autoref{fig:reach_term}.  To simplify the presentation, the code examples in
the paper use to a more common Java-like syntax.  The actual analysis is
applied to bytecode, which is closer to Carib's Jimple-like syntax of
\autoref{fig:carib}; however, the more Java-like syntax better communicates the
intuition behind our technique.  In  \autoref{fig:reach_term}, method
\lstinline+m+ has a single formal parameter \lstinline+a+ and accesses its \lstinline+b+
field and subsequently \lstinline+b+'s field \lstinline+x+ through the variable
\lstinline+tmp+.  
Thus, $RLV($\lstinline+a+$) =
\{\rep($\lstinline+a.b+$),\rep($\lstinline+tmp.x+$)\}$, which, assuming \lstinline+A+ has no super classes, 
yields \{\lstinline+A.b+, \lstinline+A.B.x+\}.

\lstset{numbers=none}
\begin{figure}[t]
\begin{center}	
\begin{tabular}{c@{\hspace{0.3in}}c}	
\begin{lstlisting}
class A { B b;}
class B {int x;}	 
\end{lstlisting}
 	& 
\begin{lstlisting}
void m(A a)
{
   a.b = ...;
   B tmp = a.b;
   tmp.x = ...;	
}
  	\end{lstlisting}
  	\\
  	&
  	\end{tabular}
  	
	\end{center}
	
	\caption{A Java example illustrating the l-values reachable from the formal parameter $a$.} 
	\label{fig:reach_term}
\end{figure}

Our transformation $\phi$ applies the rules \textsc{loop}, \textsc{rec}, and
\textsc{api}.  We naturally extend $\phi$ to an entire program $P$, where each
divergent construct $\stmt$ in $P$ is replaced with $\phi(\stmt)$ in $P'$. 

\section{\ourAnalysis: Discovering Sub-Turing Islands}
\label{sec:discovery}

This section introduces our analysis algorithm \ourAnalysis, named after the
British explorer Captain Cook.
Starting from the basic knowledge that divergent constructs are clearly part of the swamp, we want
to analyze their impact on other parts of the program.  Let us write
$\ell_1{:}\mathsf{x}$ to refer to variable \textsf{x} at program location
$\ell_1$ (e.g., at a given line number). We also write
$\mathsf{depend}(\ell_2{:}\mathsf{y},\ell_1{:}\mathsf{x}, s)$ to indicate that
in the scope of statement $s$, variable \textsf{y} at location $\ell_2$
depends on variable \textsf{x} at location $\ell_1$. In other words, modifying
\textsf{x} at $\ell_1$ may modify \textsf{y} at $\ell_2$. We over-approximate
$\semc{\cdot}$ (\autoref{fig:rules}) with respect to divergence
propagation via the following rule:
\begin{mathpar}
\inferrule* [left=(divergence propagation)]
{\ell_1{:}\mathsf{x}=\divergesymb \;\;\;\; \mathsf{depend}(\ell_2{:}\mathsf{y},\ell_1{:}\mathsf{x}, s)}{\ell_2{:}\mathsf{y}=\divergesymb}
\end{mathpar}
This suggests an approach for identifying sub-Turing islands by applying a dependency analysis whose goal is to assign $\bot$ to variables that depend on other divergence-affected variables. Broadly speaking, our analysis approximates the dependency relation induced by the program over variables, yielding an over-approximation of the swamp.  Methods not in the swamp make up the sub-Turing islands, which we thus conservatively under-approximate.

\begin{figure*}[t]
\begin{center}
\begin{tabular}{|c|}
\hline
\\
\includegraphics[scale=0.45]{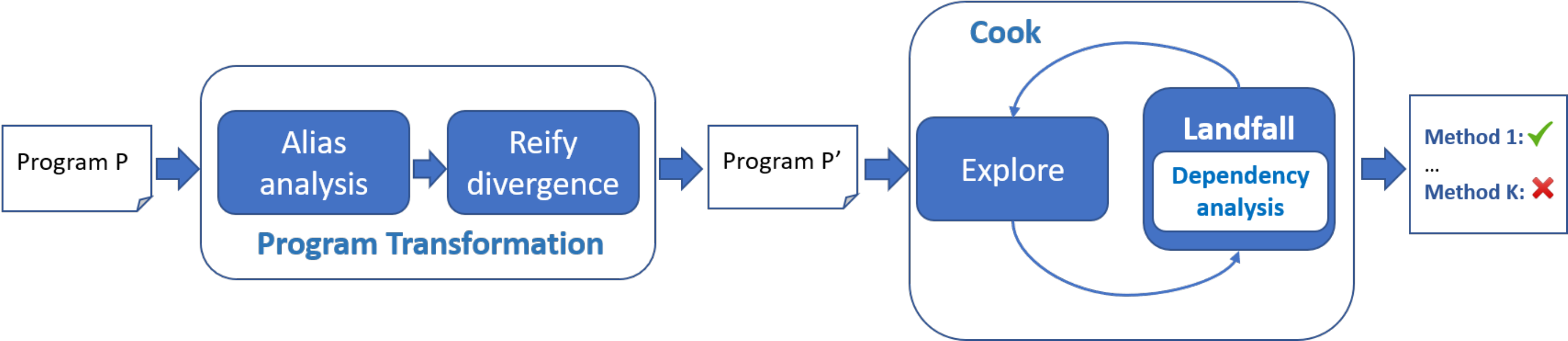}\\
\hline
\end{tabular}
\end{center}
\caption{\ourAnalysis's workflow and main components:  check marks sub-Turing
	island methods; cross marks methods in the Turing swamp.}
\label{fig:global_calypso} 
\end{figure*}

\autoref{fig:global_calypso} overviews \ourAnalysis's workflow and components.
\ourAnalysis takes as input a transformed program whose divergent constructs
have been rewritten. 
\ourAnalysis outputs a report indicating which methods are sub-Turing and which
fall into the Turing swamp.

While a sub-Turing island can be any code region, the islands we consider in the
remainder of the paper are methods.  \ourAnalysis implements a bottom-up
inter-procedural dependency analysis. It consists of two fix-point computations.
The outer computation, Explore (\autoref{alg:ExploreProgram}), operates over the whole
program and calls the inner computation Landfall (\autoref{alg:Landfall}), to compute
facts for methods.  In what follows, we describe each algorithm in detail.

\subsection{Explore: Interprocedurally Searching for Sub-Turing Islands}
\label{sec:explore} 

Starting from the transformed program, \ourAnalysis is an inter-procedural taint
analysis that propagates divergence.  \ourAnalysis assigns a method
to the swamp if it uses a tainted variable when called in a divergence-free state.
Thus, \ourAnalysis considers only taints produced by the method or a method it
transitively calls.  

In sub-Turing analysis of termination, nested loops (and recursion) can propagate bottom outwards but enclosing loops cannot propagate it inwards.
Otherwise, if non-termination were to be defined to propagate inwards, this would make the analysis of Islands often trivial and useless. 
For example, a loop-free reactive program, encased in a single non-terminating loop would often simply become `all swamp'.
That would not be helpful for analysis: the body is loop free and so this body always terminates. It can be analysed as a terminating island of code, in isolation from its surrounding loop.

In such a reactive system, figuratively speaking, the program is a single large `castle' on an island surrounded by a `moat' of swamp.
Such a `swamp castle' does not, itself, fall into the swamp.
Pragmatically, this means that we could (and we argue, should) analyse and reason about the body of such a reactive system (which is loop free) in a very different way to the way in which we would reason about it as a whole component in a larger system.
However, for our Cook analysis, the fact that taints do not propagate from the calling context means
we cannot use an off-the-shelf solution~\cite{Flowdroid}.

\ourAnalysis's output is the set of sub-Turing methods. \ourAnalysis is
inter-procedural and needs the program's call graph. Object-oriented languages,
in general, have many features, such as method overriding, that make 
constructing an exact call graph at compile time impossible.  Thus,
\ourAnalysis over-approximates the call graph using a
class hierarchy approach~\cite{SundaresanHRVLGG00} that conservatively
approximates the runtime types of receiver objects.
For an object $o$ having a declared type $C$, its estimated types will be $C$ plus all the subclasses of $C$. If $C$ is an interface then its estimated types are all the classes implementing it and the classes derived from them. We use the notation $\supseteq$ to represent the inheritance relation between classes (types); $C \supseteq S$ means that $S$ is a subclass of $C$. This relation is reflexive, thus $C \supseteq C$. Given an object $o$, the function $rt$ returns all the types that $o$ can potentially have at runtime. If the declared type of $o$ is the class $C$ then we have
\[rt(o) = \{ C' \;|\; C \supseteq C'\}.\]
Let function $Impl(I)$ return all classes implementing interface $I$, including the implementations of subinterfaces of $I$.   
If the declared type of an object $o$ is an interface $I$, then we have 
\[rt(o) = \{ C' \;|\; C \supseteq C' \wedge C \in Impl(I) \}\]
This means that we take into account all the classes implementing $I$, the ones implementing subinterfaces of $I$ and their subclasses. 
For a method invocation $o.m$, the possible resolutions of the virtual method $m$ at runtime is given by 
\[rt(o.m) = \{C.m \;|\; C \in rt(o) \wedge m \in C\}\]    
We use a class name as a prefix to distinguish different virtual methods. We write $m \in C$ to indicate that method $m$ is defined in class $C$ and use $s \in body(C.m)$ to stipulate that statement $s$ appears in method $C.m$. 
Finally, the call graph of a program $P$ is given by 
\begin{align*}
CG(P) = \{(C.m, C'.m') \;|\; C\in P \wedge m \in C \wedge o.m' \in body(C.m) \wedge\; C'.m' \in rt(o.m') \}
\end{align*}
By $C \in P$, we mean that the class $C$ is defined in the program $P$. Hence the call graph represents the set of all possible pairs of (caller, callee) belonging to the given program.
 
\begin{algorithm}[t]
\KwIn {Program $P$}
\KwOut {$P$'s sub-Turing methods}
\SetKw{Var}{Var}
\SetKw{continue}{continue}
$\mathit{worklist}$ := $M(P)$ \tcp*{$M(P)$ is the set of $P$'s methods.}
\Var map $\mathit{summary}$\;
\Var set $swamp$ := $\emptyset$\;
\ForEach{$m \in \mathit{worklist}$}
  	{
  	  $\mathit{summary}[m]$ := $\emptyset$\;
	}	
\While{$\mathit{worklist} \neq \langle\rangle$} {
  $m$ := $\mathsf{pop}(\mathit{worklist})$\;
  $s'$ := $\mathsf{Landfall}(P,m)$ \tcp*{\autoref{alg:Landfall} defines Landfall.}
  \If{$\exists (x, \bot) \in s'$}
     {
       $swamp$ := $swamp \cup \{m\}$
     }
  
  $s'$ := $s' - \mathsf{locals}(m)$ \tcp*{Remove $m$'s locals from its summary.}
  \If{$s' \neq \mathit{summary}[m]$}
     {
     	$\mathit{summary}[m]$ := $s'$\;
 	\tcc{$CG(P)$ is $P$'s call graph.}
	\ForEach{$m' \in M(P) \bullet (m',m) \in CG(P)$}
	{
	    $\mathsf{push}(\mathit{worklist}, m')$ %
  	    \tcp*{$m$'s summary changed, so we update its callers.}
	}	
     }
        
}	
\Return $M(P) - swamp$\;
\caption{Explore traverses its input program's call graph, calling Landfall on 
	each method, until it reaches a fix point where no method summaries
	change and it exhausts its worklist.}
\label{alg:ExploreProgram}
\end{algorithm}

Leveraging the approximate call graph, \autoref{alg:ExploreProgram} implements
$\mathsf{Explore}$, \ourAnalysis's interprocedural algorithm.
$\mathsf{Explore}$ takes a program transformed by $\phi$
(\autoref{sec:realising_oracles}).  
$\mathsf{Explore}$ initializes a worklist to hold
all the methods found in the program (Line 1) and associates empty summaries
with each method (Lines 4-5).  The swamp is also initially empty (Line 3).  A
summary for each method is then computed, by calling $\mathsf{Landfall}$ (Line
8), described below.  Using the facts
returned by $\mathsf{Landfall}$, Line 9 tests if $m$ belongs in the swamp.  It
does when its summary contains at least one element of the form
$(\mathsf{x},\bot)$, meaning that \ourAnalysis cannot safely, statically
determine that it terminates.  If this is the case, $\mathsf{Explore}$ adds $m$
to the $\mathit{swamp}$.  Function \textsf{locals} returns the set of l-value
representatives corresponding to a method's local variables. It is useless to
keep such elements in a summary; Line 11 discards them. If $m$'s summary has
changed (Line 14), its entry is updated and its callers are placed on the
worklist (Lines 15).  Finally, on Line 16 $\mathsf{Explore}$ returns the set of sub-Turing
methods as the complement of $\mathit{swamp}$ against the set of all program methods.

\subsection{Landfall: \ourAnalysis's Intraprocedural Analysis}

\textsf{Landfall} (\autoref{alg:Landfall}) is an intra-procedural analysis.  It
approximates the dependence relation induced by a given method $m$ over program
variables. It uses the lifted set of l-values $L_\bot = L \cup \{\bot\}$
and an abstract interpretation over the domain $D$ representing the powerset of
pairs of l-value representatives: 
\[
D  = \mathbb{P}(\{(\rep(\mathsf{x}),\rep(\mathsf{y})) \;|\; \mathsf{x},\mathsf{y} \in L_\bot\}),
\]
\noindent where $\rep(\bot)$ is defined as $\bot$.  Each pair
$(\mathsf{x},\mathsf{y})$ in $D$ means that \textsf{x} depends on \textsf{y}
with the use of $\rep$ taking aliasing into account.  We call the pair
$(\mathsf{x},\mathsf{y})$ a \emph{fact}. Furthermore, the element
$(\mathsf{x},\bot)$ expresses that we cannot rule out the possibility that
\textsf{x} might be affected by divergence.  

\textsf{Landfall} computes the
transitive closure over elements from the domain $D$ with respect to statements
of method $m$ using two auxiliary functions: control-dependence function
$\mathsf{control\_dep}$ and data-dependence function $\mathsf{data\_dep}$. 
The function $\mathsf{control\_dep}$ captures control dependencies created by
conditional statements. For example, consider the code shown in
\autoref{fig:implicit_dep}.  If in this example we only account for data
dependencies, we conclude that variable \textsf{y} only depends on \textsf{z}
errantly omitting \textsf{x}.  However, if \textsf{x} is affected by divergence,
we need to propagate this fact to \textsf{y}. 

Before describing how we compute $\mathsf{control\_dep}$, we introduce relevant terminology. Each method in the
program is represented by a Control Flow Graph (CFG), a directed graph $(N,E)$
where $N$ is the set of nodes and $E$ a set of edges.  Each node represents
either an assignment or a branch condition.  The edges, ${E \subseteq N
\times N}$, represent control flow between program statements.  In Carib, we map
each assignment to a node with one successor and each conditional statement to a
node with two successors, representing the $\true$ and $\false$ branches.  For
CFG node $n$, $\mysucc(n)$ is the set of successors of $n$, $\mathsf{pred}(n)$
its predecessors, and $\mathsf{stmt}(n)$ the statement $n$ represents.  Finally,
each CFG includes two special nodes: $\mathsf{entry}(\mathit{CFG})$ is the CFG's
unique entry node, which has no predecessors, and $\mathsf{exit}(\mathit{CFG})$
is its unique exit node, which has no successors.  

To compute control dependencies, we use the well-established approach of
Ferrante \etal~\cite{FerranteOW87}, which we denote as
$\mathsf{control\_dep}(\mathit{CFG}, \ell , M)$ where \emph{CFG} is a control
flow graph, $\ell$ a location, and $M$ a map associating locations with sets of
facts.  This function returns the set of facts induced by control dependencies
for location $\ell$. Function $\mathsf{control\_dep}$ includes transitive
control dependencies. 

\lstset{numbers=left}
\begin{figure}
\begin{center}	
\begin{tabular}{c}	
\hline
\begin{lstlisting}
	y = 0;
	if (x > 0)
	    y = z;
\end{lstlisting}
\\
\hline
\end{tabular}
\end{center}
\caption{Code illustrating a case of control (implicit) dependency. Variable \textsf{y} is control-dependent on variable \textsf{x}.} 
\label{fig:implicit_dep}
\end{figure}

Turning to the data dependences, 
the function ${\mathsf{data\_dep}: D\times L(\mathcal{G})\longrightarrow D}$
models the effect of program statements on elements of the abstract domain $D$.
For a given fact $d \in D$ and statement $\stmt \in L(\mathcal{G})$,
$\mathsf{data\_dep}$ is defined as follows: 
\begin{align*}
\mathsf{data\_dep}(d, \stmt) = & \{ (x,y) \;|\; \exists z.\; (x,z) \in \mathsf{gen}(\stmt) \wedge (z,y) \in d\} 
\; \cup\;  (d - \mathsf{kill}(\stmt, d)).
\end{align*}

\noindent
where $\mathsf{gen} (\stmt)$ is the set of dependencies locally induced by statement $\stmt$.
For example, $\mathsf{gen}(\mathsf{x:= y + z})$ yields $\{\mathsf{(x,y)}, \mathsf{(x,z)}\}$. 
Function \textsf{data\_dep} transitively extends the relation represented by the input facts and the relation induced by the $\stmt$.  It also excludes (kills) facts that are no longer valid after the assignment. For example,
\[
\mathsf{data\_dep}(\{(\mathsf{x},\mathsf{t}), (\mathsf{y},\mathsf{p})\}, \mathsf{x}:= \mathsf{y}) = \{(\mathsf{x},\mathsf{p}), (\mathsf{y},\mathsf{p})\}.
\]          
Since the assignment modifies \textsf{x}, the fact $(\mathsf{x},\mathsf{t})$ no
longer holds. Landfall transitively obtains the fact $(\mathsf{x},\mathsf{p})$
from the input fact $(\mathsf{y},\mathsf{p})$ combined with
$(\mathsf{x},\mathsf{y})$ from the assignment statement. 

We provide the definitions of functions $\mathsf{gen}$ and $\mathsf{kill}$ for
Carib's basic statements in \autoref{tab:dep_kill}.  Assignments to simple
variables (the first five cases) result in dependencies expressing how the
assignment's left-hand-side depends on the identifiers appearing in its
right-hand-side except when the right-hand-side is a constant, which does not
introduce any dependencies.  When the right-hand-side is an object field or an
array reference, we use its representative to take aliases into account. The
return statement $\mathsf{return}\;id$ is modeled as the assignment \textsf{ret
:= id}, where \textsf{ret} is a special variable (see \autoref{fig:rules}) used
to store and retrieve the method's return value.
In all these cases, we kill
input facts expressing dependencies involving the assignment's left-hand-side. 

In case of an assignment to a field or array element, we use its representative
to take aliases into account. To preserve soundness, we do not kill any facts.
Indeed, a representative over-approximates possible aliases.  Therefore, the
updated l-values may or may not be an actual alias of a given fact.  For a call
to a method $m$, we replace the formal parameters with the corresponding actuals
in $m$'s summary, which is a set of facts expressing dependencies induced by
$m$.  We also replace the special variable \textsf{ret} with \textsf{r}.
Landfall computes method summaries iteratively, on-the-fly when demanded by
\textsf{Explore}.  Finally, for the
assignment $id := \bot$, we keep the fact expressing that the assigned variable
is affected by divergence because the purpose of our analysis is to track the
propagation of $\bot$. 

\begin{table*}[t]
\begin{center}
\begin{tabular}{lcc}
\hline
 Statement $\stmt$ & $\mathsf{gen}(\stmt)$ & $\mathsf{kill}(\stmt, d)$\\
\hline
 $id := c$ & $\emptyset$ & $\{(x,y) \in d \;|\; x = id\}$\\

 $id_1 := id_2$ & $\{(id_1, id_2)\}$& $\{(x,y) \in d \;|\; x = id_1\}$\\

$id_1 := \mathsf{op}\;id_2$ & $\{(id_1,id_2)\}$&  $\{(x,y) \in d \;|\; x = id_1\}$\\

$id_1 := id_2 \;\mathsf{op}\;id_3$ & $\{(id_1, id_2), (id_1, id_3)\}$& $\{(x,y) \in d \;|\; x = id_1\}$\\
$id_1 := id_2[id_3]$ & $\{(id_1, \rep(id_2[id_3]))\}$&  $\{(x,y) \in d \;|\; x = id_1\}$\\
$id_1[id_2] := id_3$ & $\{(\rep(id_1[id_2]), id_3)\}$&  $\emptyset$\\
$\mathsf{return}\;id$ & $\{(ret,id)\}$ & $ \emptyset$\\
$r := m(\vec{Y})$ & $\mathit{summary}(m)[\vec{Y}/\vec{X}, r/ret]$& $d - \{(x,y) \in d \;|\; x = r\}$\\
 $id := \bot$ & $\{(id,\bot)\}$ & $\{(x,y) \in d \;|\; x = id\}$\\
\hline
\end{tabular}
\end{center}
\caption{Definition of $\mathsf{gen}$ and $\mathsf{kill}$ for 
	relevant Carib statements; for call statements, $\vec{X}$ 
	contains $m$'s formals.}
\label{tab:dep_kill}
\end{table*}

Landfall uses $\mathsf{control\_dep}$ and $\mathsf{data\_dep}$ in a standard
worklist.  The input and output
of all nodes is initialized the empty set on Lines 4-5.  Then, the entry node's
input is created on Line 6.  New facts are produced by simulating the effect of
program statements using the transfer function \textsf{data\_dep} (Line 12),
accounting for control dependencies (Line 13). When the set of facts associated
with a given location $n$ changes, all successors of $n$ are explored again
(Lines 14-16). The algorithm is guaranteed to terminate because $L_\bot$ is
finite and so is the set of facts. Once a fix-point is reached, the algorithm
returns the set of facts accumulated at the exit node. 
   
\begin{algorithm}[t]
\KwIn {Program $P$, method $m$}
\KwOut {set of facts}
\SetKw{Var}{Var}
\SetKw{continue}{continue}
\Var map $\mathit{IN}$, $\mathit{OUT}$\;
Let $\mathit{CFG}$ be the control flow graph of $m$\;
Let $\mathit{L_\bot}$ be the lifted set of l-values appearing in $P$\;
\ForEach{$n \in \mathsf{node}(\mathit{CFG}) $}
  	{
  		$\mathit{IN}[n]$ := $\mathit{OUT}[n]$ := $\emptyset$\;	
	}	
$\mathit{IN}[\mathit{entry}(\mathit{CFG})] = \{(\rep(x),\rep(x))\;|\; x \in \mathit{L_\bot}\}$\;

$\mathit{worklist}$ := $\mathsf{push}(\mathit{entry}(CFG))$\;
\While{$\mathit{worklist} \neq \langle\rangle$}
{
  $n$ := $\mathsf{pop}(\mathit{worklist})$\;
  $\mathit{OUT_0}$ := $\mathit{OUT}[n]$\;
  $\mathit{IN}[n]$ := $\bigcup_{n' \in \mathsf{pred}(n)}OUT[n']$\;
  $\mathit{OUT}[n]$ := $\mathsf{data\_dep}(\mathit{IN}[n],\mathsf{stmt}(n))$\;
  $\mathit{OUT}[n]$ := $\mathit{OUT}[n] \cup \mathsf{control\_dep}(\mathit{CFG}, n, \mathit{IN})$\;
  \If{$\mathit{OUT_0} \neq \mathit{OUT}[n]$}
     {
     	\ForEach{$n' \in \mathsf{succ}(n)$}
			{
			 	$\mathsf{push}(\mathit{worklist}, n')$\;
			}	
     }
}	
\Return $\mathit{OUT}[\mathsf{exit}(\mathit{CFG})]$\;
\caption{Landfall approximates the dependency relation over program variables that 
	its input method defines; \autoref{tab:dep_kill} defines its
	$\mathsf{gen}$ and $\mathsf{kill}$ functions..}
\label{alg:Landfall}
\end{algorithm}

\subsection{Implementation}

We implemented our approach for sub-Turing island identification in a tool
called \ourTool, which is written in Python. \ourTool takes as input an Android
application and returns a report that includes the analysis result together with
other statistics. \ourTool accepts Android apps directly in binary (APK) format.
It uses Androguard\footnote{https://github.com/androguard/androguard} to parse
and decompile the APK files as well as generate the control flow graphs.  Hence,
\ourTool does not require source code.  We use our own intermediary
representation for instructions which has a lisp-like format. One key phase in
\ourTool is loop extraction (\autoref{sec:loops}), which extracts  a list of
loops, each of which is identified by its header together with the nodes it
contains. It also obtains the hierarchical (domination) relation between loops.
Finally, \ourTool implements the over-approximation of the call graph based on
the class hierarchy approach \cite{SundaresanHRVLGG00} (\autoref{sec:explore}).   
Endeavour is available at \url{to.be@posted.post.final.acceptance}.


\section{Experimental Results}
\label{sec:experiments}
This section empirically investigates six research questions involving sub-Turing islands, henceforth abbreviated ST-islands.
We start by overviewing the application corpora that makes up our experimental subjects.    
The investigation then begins by considering the prevalence of ST-islands.
Simply put if ST-islands are rare then their study is of little practical value.
We next take a deeper look in into the main causes of divergence.
Finding API methods the dominant source, we consider the impact of \emph{safe
listing} subsets of the API methods.
Then turning to two of the many applications of ST-islands, we consider
first the relationship between bug density in the swamp and on the ST-islands,
and second the percentage of verification conditions, such array bound
violations and null object dereferences that occur on ST-islands.
Finally. we consider the runtime efficiency of our tool \ourTool.


In the experiments, unless otherwise stated, we make the following assumptions.
First, we discarded getters and setters as we assume that they are implemented in a standard way making them trivially sub-Turing. 
In addition, we initially assume that {\bf all} API calls diverge and bind $\bot$ to all variables they may write or that depend on them.

We study two sets of apps. A large dataset, \textsc{app\_bin}, of over one thousand apps, for which source code is unavailable, and a smaller set, \textsc{app\_src}, of ten apps, for which full source code is available. 
Both corpora, are composed of a range of real world production apps to ensure that our empirical scientific findings have high external validity. 
%
The \textsc{app\_bin} dataset is composed of 1100 Android applications uniformly selected from more than 600$\,$000 apps collected from the Androzoo\footnote{https://androzoo.uni.lu}. Androzoo apps have diverse origins, including the Google Play, store which is the predominant source the apps we study. Our set of 1100 apps contains more than 2 million methods. 
The \textsc{app\_src} dataset is composed of ten applications selected from Github under certain criteria that we describe later. We only consider this dataset in the experiment described in \autoref{sec:bug_dist}, which requires the app source code. In all other experiments, we consider the larger \textsc{add\_bin} set.

\subsection{Landscape of ST-islands}
\label{sec:st-landscape}
First of all it is important to know the proportion of code that resides within ST-islands. The answer to this question suggests the code size over which we can reason precisely. 
A significant proportion means that it is worth investing in the improvement of static analysis as the benefit may be substantial. 
So the first research question we address is the following:

\vspace{2mm}
\noindent\textbf{RQ1}: \emph{What is the proportion of code occupied by ST-islands?}\vspace{-1mm}
\vspace{2mm}

\begin{figure}[!ht]
\begin{center}
\begin{tabular}{@{\hspace{-.1in}}c}
\includegraphics[scale=0.4]{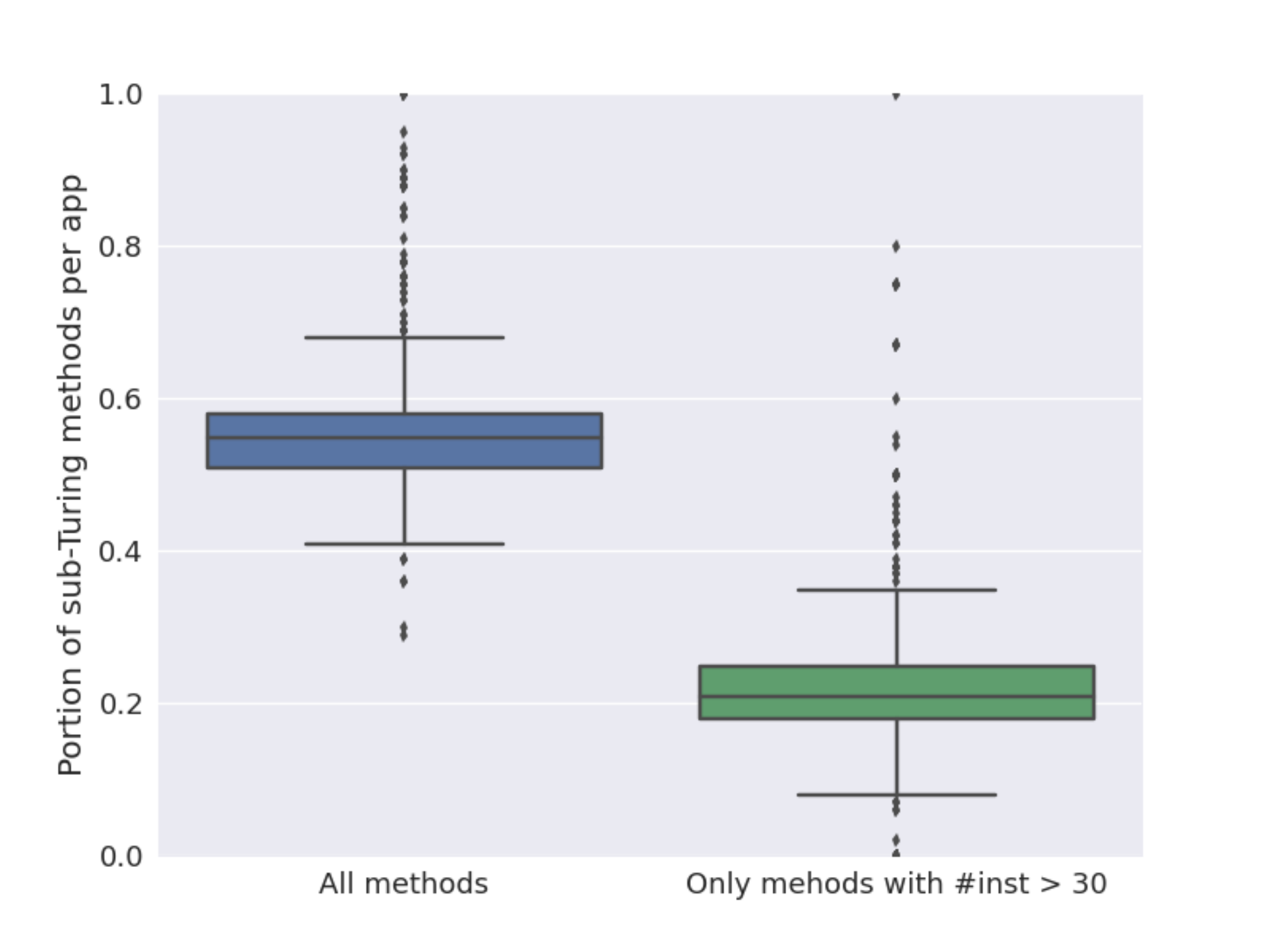}
\\
\end{tabular}
\end{center}
\caption{Percentage distribution of ST-methods in our 1100 apps, discarding getters and setters (left boxplot). In addition to discarding getters and setters we also discard methods with less than 30 bytecode instructions (right boxplot). The average percentage of ST-methods in the first case (left) is 55\% and it is 22\% for the second case (right).}
\label{fig:space_in_code} 
\end{figure}

The results using \textsc{app\_bin} are summarized in \autoref{fig:space_in_code} 
where the left boxplot shows the distribution of ST-method percentages. 
\begin{framed}
\vspace{-1mm}
\noindent\textbf{Finding 1a}: Overall, the average percentage of ST-methods in an app is approximately 55\%, hence, the majority of methods are sub-Turing. 
\end{framed}

To study the impact of code size on our results, we want to exclude trivial methods. Defining trivial is hard. 
We conservatively consider methods of fewer than ten lines as trivial.  To convert lines into bytecode instructions, we averaged method length in bytecode instructions over its non-comment source code length and found that on average
each line of source code generates three bytecode instruction. 
Thus, we consider a method trivial if it includes fewer than thirty bytecode instructions. 
The results when considering only non-trivial methods are shown on the right of \autoref{fig:space_in_code}. 
\begin{framed}
\vspace{-1mm}

\noindent\textbf{Finding 1b}: 
Discounting trivial methods, the percentage of ST-methods is 22\%, which while lower than overall average, still represents a significant portion of the code.
\end{framed}
While the percentage of ST-methods drops, it remains significant as it represents almost a quarter of each app. Moreover, our analysis is both sound and efficient, hence the percentage under estimates the true proportion of code that lies in non-trivial sub-Turing islands. A more precise but less efficient analysis can only ever uncover additional sub-Turing methods.
Hence, this result underscores the value of investing in static analysis tools specialized to exploit ST-islands. 

\subsection{Causes of Divergence}  
\label{sec:div_cause}
Understanding the causes of divergence informs us about prevalent reasons of precision loss. For example, if it turns out that a certain language construct is the dominant cause of divergence, then we might want to give it greater attention in future work. Therefore, we seek an answer to the following research question: 

\vspace{2mm}
\noindent\textbf{RQ2}: \emph{What are the main causes of divergence?} 
\vspace{2mm}

To answer this question, we refined our analysis by extending the abstract domain with an element indicating the cause of divergence: API call, loop, or recursive method. 

\begin{framed}
\vspace{-1mm}
\noindent\textbf{Finding 2}: 
Over corpus $\textsc{app\_bin}$, we classify the sources of divergence as following
\begin{center}
\begin{tabular}{ l c c c} 
\textsf{\bf api} & \textsf{ \bf loop} & \textsf{\bf recursion} \\
{\bf 76\%} & {\bf 13\%} & {\bf 11\%} \\
\end{tabular} 
\end{center}
\vspace{-1mm}
\end{framed}

\noindent
We can see that over three quarters of the divergence is due to library API calls.
This suggests that a more precise modelling of API calls is likely to improve the precision of a given static analysis. We set out to experimentally investigate this hypothesis in the next section.

\subsection{API Safe Listing}
\label{sec:api_list}

\begin{figure*}[!ht]
\begin{center}
\begin{tabular}{@{\hspace{-.3in}}c@{\hspace{-.3in}}c}
\includegraphics[scale=0.4]{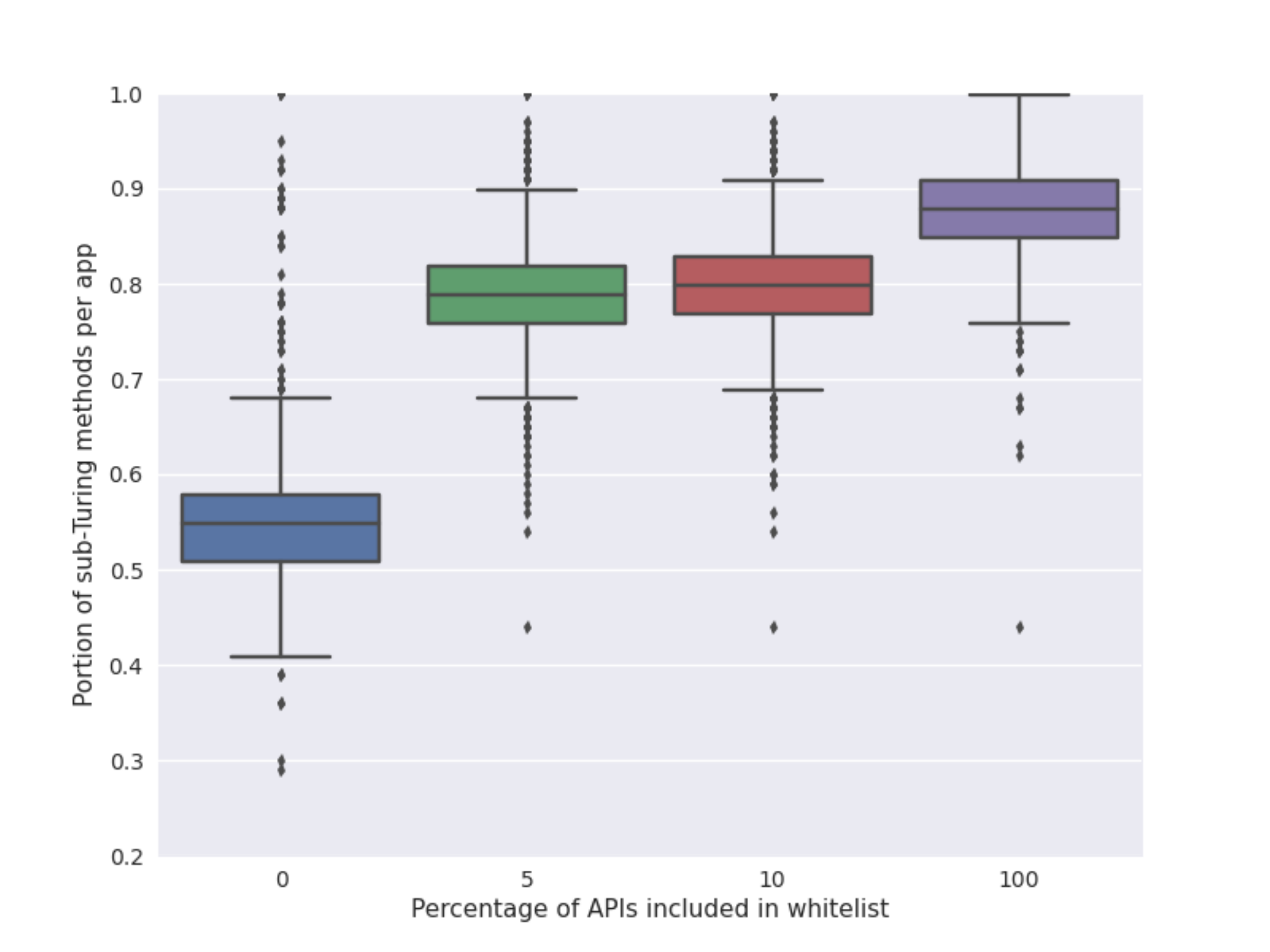}
&
\includegraphics[scale=0.4]{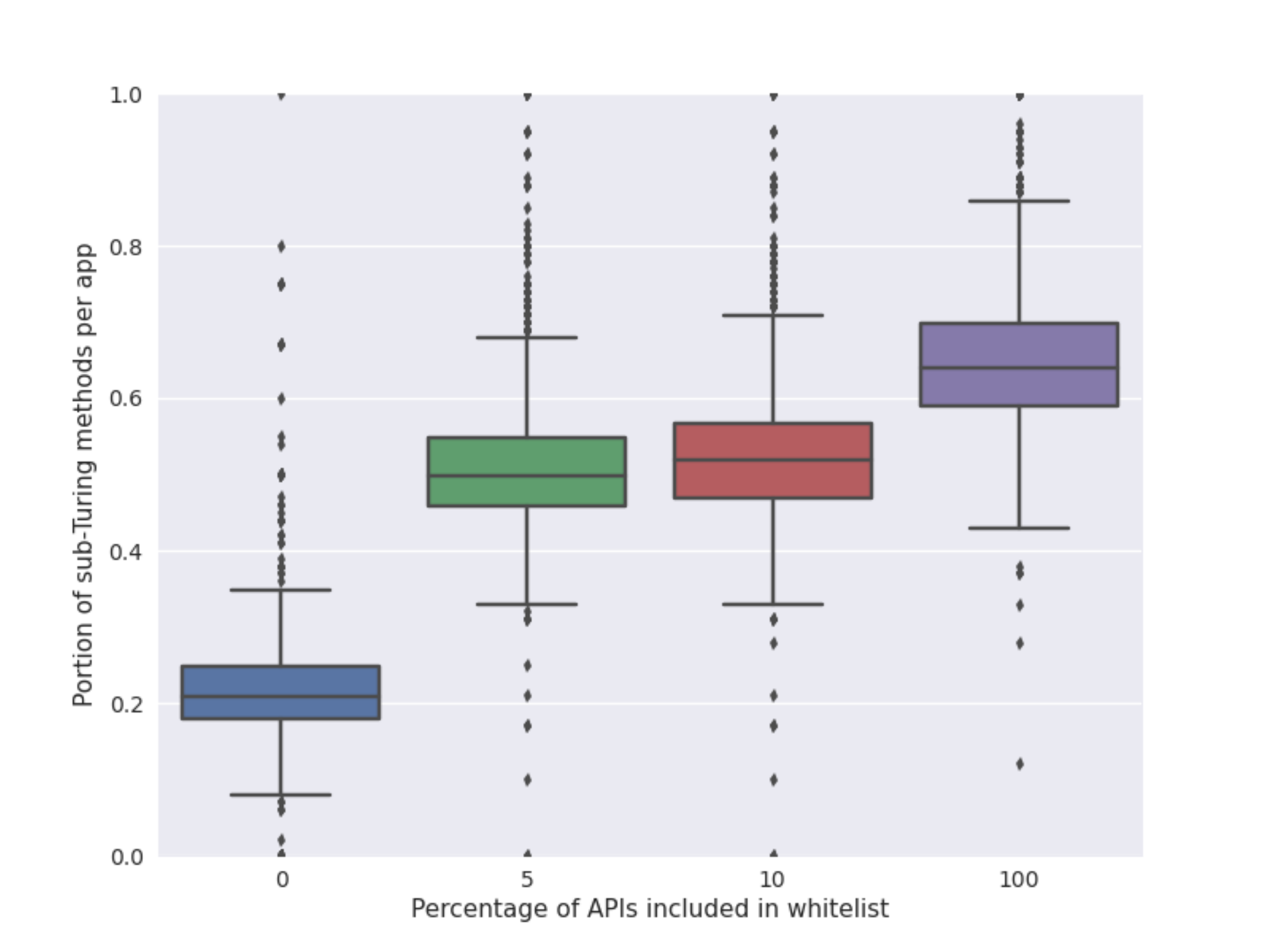}
\\
(a) : All methods.
&
(b) : Only methods with \#inst $>$ 30.
\\
\end{tabular}
\end{center}
\caption{Percentage distribution of ST-methods considering a safe list of most frequently used APIs. 
The $x$-axis shows the size of the safe list as a percentage of most frequently used APIs.
Chart (a) shows box plots for all 1100 apps, discarding getters and setters while Chart(b)
also discards methods with fewer than 30 bytecode instructions.}
\label{fig:api_whitelist} 
\end{figure*}

\ourAnalysis is very conservative as it assumes that all API calls cause divergence. In practice, many called API methods have a quite well-understood and documented behaviour, making it is plausible to assume that calls to such API methods are not a source of divergence. In this section, we test the impact of this possibility in the following research question: 

\vspace{2mm}
\noindent\textbf{RQ3}: 
\emph{How does a more precise modelling of APIs impact the analysis?} 
\vspace{2mm}

We define a safe list of most frequently used APIs which are assumed to not induce divergence. Among the selected APIs are methods from the Java standard library and some Android frequently used API methods. Under this setting, we repeat the experiments of \autoref{sec:st-landscape}, where we vary the size of the API safe list. Results are shown in \autoref{fig:api_whitelist}.  \autoref{fig:api_whitelist}a, shows the percentage of ST-methods per app for different sizes of the API safe list while \autoref{fig:api_whitelist}b considers only methods with more than 30 bytecode instructions. 
Results when using an empty API safe list repeat the data shown in \autoref{fig:space_in_code}. 
We included them as a baseline. 
At the other end placing all API methods on the safe list allows us to investigate the impact of a developer who seeks to focus the analysis solely on his or her code.

\begin{framed}
\vspace{-1mm}
\noindent\textbf{Finding 3}: 
For a safe list containing just 5\% of most frequently used APIs, the average percentage of ST-methods grows to almost 80\% when all methods are considered and just over 50\% when only methods containing more than 30 bytecode are considered. 
\vspace{-1mm}
\end{framed}          

%
\noindent
Here a safe list of only 5\% of the frequently used APIs yields an important increase in ST-methods.
Interestingly, increasing this to 10\% has minimal impact, which may be an instance of the way the most frequently used calls tend to distribute as a power law.
Finally, including all APIs on the safe list causes 88\% of all methods and 66\% of all non-trivial methods to be ST-methods.
The trend here hints at the value in techniques such as providing formal summaries for the common API methods. 

\subsection{Distribution of Bugs over ST-Islands}
\label{sec:bug_dist}
It is interesting to check whether there is a correlation between bugs and ST-islands. We address this possibility in the following research question: 

\vspace{2mm}
\noindent\textbf{RQ4}: \emph{Is there a significant difference in the bug distribution in the swamp compared to the ST-islands?} 
\vspace{2mm}

Investigation of this research question requires application source code;
thus we make use the \textsc{app\_src} collection, which was collected under the following constraints:
\begin{itemize}
\item {\bf Open source:} we need the code of the application as well as the corresponding repository to perform the experiment.
\item {\bf Repository history:} to rule out simple weekend projects.
\item {\bf Non-trivial size:} to rule out small toy applications. 
\item {\bf Number of application installations:} we want the apps to have real users, thereby attesting to their practical use.    
\end{itemize}    
\noindent

The resulting \textsc{app\_src} collection includes the ten real-world applications shown in \autoref{tab:bug_dense}. 

We compute bug density for ST-methods and swamp methods using the following steps:

\begin{itemize}
\item To identify bugs and their corresponding locations, we use a heuristic based on a bag of words. We check the presence of certain commits  associated with keywords such as "bug", "fix", etc.~in the \texttt{git} repository of each application. We call such commits \emph{bug-fix} commits. 

\item A buggy line is any line removed, added, or modified by a bug-fixing commit.  A method is buggy if it contains a buggy line. We assume that a single bug is associated with a single commit and write bugs(m) to express the number of bugs associated with method m.

\item As our analysis is at the bytecode level, we compile the original source code of each app considered to obtain a binary APK file to analyse.  

\item Finally, we compute bug density for ST-methods and swamp methods. The bug density for an application $A$, $\bugDens(A)$, is defined as 
\[
  \bugDens(A) = \frac{1}{|A|}\sum_{m \in A} \frac{\mathsf{bugs}(m)}{\mathsf{LoC}(m)}
\]

where $\mathsf{LoC}(m)$ is the number of lines of code in method $m$ and $|A|$ the number of methods in $A$.
We respectively denote the bug density for ST-methods and swamp methods as $\bugDens_{ST}(A)$ and $\bugDens_{SW}(A)$.
\end{itemize}

\begin{table}
\begin{center}
\begin{tabular}{ l r S[table-format=2.1] S[table-format=2.1]} 
\hline
\multicolumn{1}{c}{app $A$} & \multicolumn{1}{c}{LOC} & \multicolumn{1}{c}{$\bugDens_{ST}(A)$}  & \multicolumn{1}{c}{$\bugDens_{SW}(A)$}\\
\hline
bitcoinwallet &  23$\,$392 &  0.5  &   1.0 \\
connectbot    &  26$\,$625 &  0.0  &  18.4 \\
irccloud      &  57$\,$471 &  5.7  &  12.0 \\
k9            & 123$\,$606 & 55.5  &  59.3 \\
mgit          &  10$\,$919 & 11.7  &  13.1 \\
orbot         &  18$\,$772 &  0.0  &   1.1 \\
owncloud      &  63$\,$495 & 48.1  &  57.2 \\
signal        &  92$\,$868 & 29.2  &  18.8 \\
vlc           &  69$\,$976 &  8.6  &  10.7 \\
worldpress    & 128$\,$433 & 24.1  &  22.7 \\

\hline
\end{tabular} 
\end{center}
\caption{Bug Density in ST-methods and the swamp for 10 Android open source projects given as number of bugs per kilo line.}
\label{tab:bug_dense}
\end{table}

Overall in \textsc{app\_src} there are 6906 ST-methods comprised of 475 KLoC with 1863 bugs,
and 7417 swamp methods comprised of 894 KLoC with 5317 bugs.
We compare \emph{bugginess} statistically using the non-parametric Wilcoxon
test at first the method level and then the line level.
The average bugs per method of 0.27 for ST-methods and 0.72 for the swamp
are statistically different ($p < 0.0001$).
Because swamp methods tend to include more lines of code, we also compare the
two using bugs-per-line.
In this case, the 0.0265 for ST-methods is again statistically less than 
the 0.289 for the swamp ($p < 0.0001$).
\autoref{tab:bug_dense} breaks these bug density out by program.

Finally, we use generalized linear models to investigate the question ``How
likely is a method to be buggy?'' where a method is considered buggy if it
contains one or more bugs.
A method's bugginess forms each model's response variable.
Generalized linear models enable us to consider multiple explanatory variables
as well as binary response variables.
In the first model, we use ST-island as the sole explanatory variable.
With an odd ratio of 2.07, the model predicts that a swamp method is over twice
as likely to contain a bug when compare to an ST-island method ($p < 0.0001$).
Including program as an additional explanatory variable, which enables the
model to account for differences between programs, increases the odds ratio
to 2.09.
The impact of additionally including lines as an explanatory variable is
negligible with or without the program variable.
Finally, it is interesting that there is no significant
interaction between program and a method being an ST-method;
thus, the likelihood of being an ST-island method is independent of the program.
This unexpected uniformity strongly supports the external validity of our findings.

%


\begin{framed}
\vspace{-1mm}
\noindent\textbf{Finding 4}: 
The bug densities for ST-islands are statistically smaller than that of the swamp ($p < 0.001$).
\vspace{-1mm}
\end{framed}

\noindent
From the above statistics, bug density tends to be higher in the swamp. This result further supports our suggestion to use the swamp as a hint for guiding bug search.
In other words, one should allocated a limited budget (time, resources, etc.) to the swamp than to the ST-islands. 

\subsection{Finding Potential Errors}
\label{sec:runtime_errors}
ST-islands are portions of code about which we can precisely answer whether a given property holds. We would like to investigate the presence of concrete properties falling into ST-islands on which program safety relies. One such property is a runtime exception such as an array out-of-bounds and null-object dereferences. We address the following research question:

\vspace{2mm}
\noindent\textbf{RQ5}: \emph{What is the percentage of verification conditions related to detecting bound violations and null object dereference runtime errors that occur in ST-islands?}   
\vspace{2mm}

We studied the spread of these two potential runtime exceptions over ST-islands in our \textsc{app\_bin} corpus of 1100 applications. We count all array accesses and object dereferences in the code and compute the proportion of the ones occurring in ST-methods for each application. 

\begin{figure}[!ht]
\includegraphics[scale=0.4]{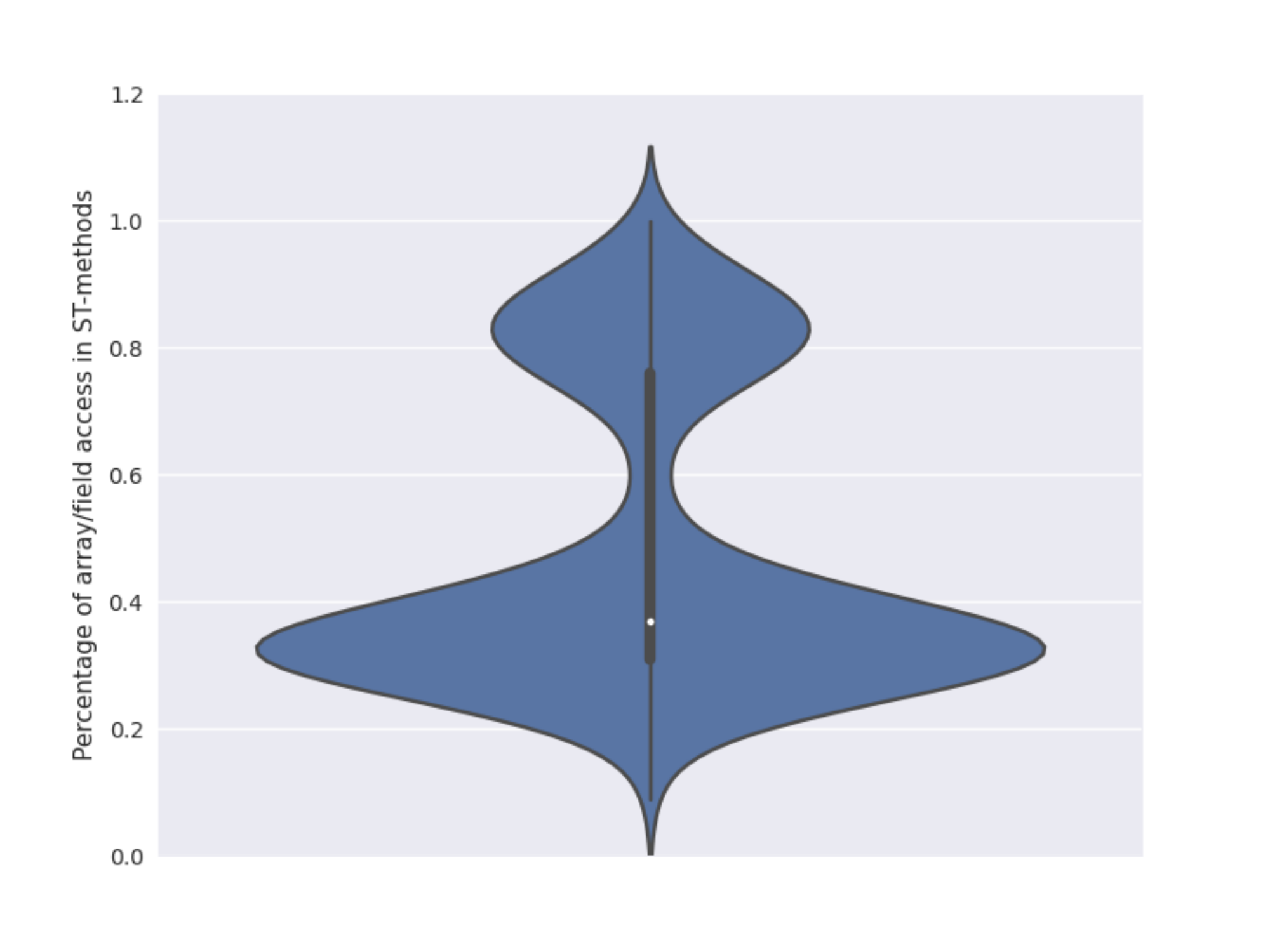}
\caption{Percentage distribution of array accesses and object dereferences in
ST-methods per application.   The $x$-axis depicts a kernel density plot of the
data, mirrored around the plot's central line.  Intuitively, kernel density
captures the likelihood that the $y$-axis has this value.}
\label{fig:rexcept_in_islands} 
\end{figure}

The results, presented in \autoref{fig:rexcept_in_islands}, show that just over one in three exceptions can be precisely checked at compile time because it lies on a sub-Turing island. This is a lower bound for our corpus of 1100 apps, because our determination of sub-Turing islands is a safe under-approximation. Moreover, as visible in the violin plot (\autoref{fig:rexcept_in_islands}), the percentage of array accesses and object dereferences is around 80\% for a notable number of apps.

\begin{framed}
\vspace{-1mm}
\noindent\textbf{Finding 6}: 
A lower bound on the average percentage of sub-Turing array accesses and object dereferences in our corpus is 37\%.  
\vspace{-1mm}
\end{framed}

\subsection{Analysis Performance}
\label{sec:analysis_perform}
We have established that non-trivial portions of real world Android app code lie in sub-Turing islands and have demonstrated that this has implications for bug density and verification in an empirical analysis. Finally, we report on the computational cost of identifying sub-Turing islands using our approximation. While many other techniques for approximation could be used, and should be explored in future work, it is useful to know whether, at least {\em one} such analysis exists that is scalable.
If we are able to provide evidence that our approximation is computationally feasible and, therefore, that there does exist a scalable useful approximation to sub-Turing islands, this will further underscore the practical value of sub-Turing analysis.

\vspace{2mm}
\noindent\textbf{RQ6}: \emph{Can ST-islands be efficiently identified?} 
\vspace{2mm}

We measured \ourTool's analysis time from parsing an application to delivering its output on a 3.2GHz Intel Core i5 quad-core processor with 8GB of memory, running Linux. 
The results show that our approach is scalable to real-world applications.

\begin{framed}
\vspace{-1mm}
\noindent\textbf{Finding 7}: \ourTool takes less than four minutes for even the largest applications studied, containing more than $40\;000$ methods. \vspace{-1mm}
\end{framed}

\section{Related Work}
\label{sec:related_work}
Our analysis marries taint analysis with termination reification (as divergence). Taint analysis is a technique used in software security~\cite{Flowdroid, Enck:2014, WeiROR14, TrippPCCG13, GordonKPGNR15}. The goal of taint analysis is to show the absence of information leaks from a set of given sources to a set of given sinks. It can be performed statically~\cite{Flowdroid, WeiROR14, TrippPCCG13, GordonKPGNR15} or dynamically~~\cite{Enck:2014}. Our bottom-up inter-procedural data-flow analysis is a flow-sensitive taint analysis that takes into account implicit information flows due to control dependencies. In our case, sources are divergent constructs.  
Our work also relates to various other topics, including invariant generation, loop summarization, bounded model checking, termination analysis, strictness analysis and program slicing. 

\subsection{Loops} 
As loops are a key component in our study, we consider work from the literature aimed at their analysis.

\paragraph{\bf Summaries.} 
Our modelling of potentially non-terminating loops consists of assigning divergence values to variables they possibly modify. Loop summarization techniques allow to infer loop-free code that soundly approximates a given loop.
 
Sharygina and Browne proposed a syntactic transformation for abstracting branches 
in loops in a UML dialect (design level)~\cite{SharyginaB03}. Kroening and Weissenbacher proposed an approach based on associating recurrence equations with loop variables and then computing a closed form for each equation. Kroening et al.~\cite{KroeningSTTW08} a proposed related technique for replacing code fragments, including loops, with corresponding abstract transformers that play the role of the summaries. Seghir proposed a lightweight technique for inferring loop summaries over array segments as well as simple variables using a set of inference rules~\cite{Seghir11}. Xie et al. presented a technique for summarizing loops that contain multiple paths and manipulate strings, with conditions over string content~\cite{XieLLLC15}. They further extended their work to support disjunctive reasoning~\cite{XieCLLL16}.
Loop summarization can be folded into our approach to increase the number of loops that can be statically determined to terminate by construction.

\paragraph{\bf Invariants} 
One approach for reasoning about loops in the context of program verification is through loop invariants~\cite{Hoare69}. Many verification tools rely on manually provided invariants~\cite{FlanaganLLNSS02, BarnettCDJL05, DahlweidMSTS09}. However, the literature is rich in terms of approaches that automatically infer invariants in various domains: arithmetic~\cite{Karr76, Muller-OlmS04, ColonSS03} (linear),~\cite{SankaranarayananSM04} (non-linear), arrays~\cite{JhalaM07,GulwaniMT08,SrivastavaG09} and heaps~\cite{SagivRW02, PodelskiW05}. 
Software model checkers attempt to build invariants automatically, during the verification process~\cite{BallRaj, HenzingerJMS02, ChakiETAL03, PodelskiRybalARMC, IvanicicSGG05, cksy2004}, relying on a popular technique called predicate abstraction~\cite{GrafS97}. 
We can use invariants to express state changes (transitions) by introducing fresh variables to symbolically model initial values of variables. Hence, similar to summaries, we can use them to express the effect of a given loop, which should improve our algorithm's precision. 
 
\paragraph{\bf Termination}

Termination is another issue related to loops. Knowing the after-state of a given loop is only possible when the loop terminates. Therefore, we model the effect of potentially non-terminating loops by assigning a divergent value to potentially modified variables. The literature is rich with work regarding termination analysis~\cite{PodelskiRybalARMC,PodelskiR04, UrbanGK16, PodelskiR05, PodelskiR04LICS, CookPR05}. So-called ranking functions~\cite{PodelskiR04, UrbanGK16} and transition invariants~\cite{PodelskiRybalARMC,PodelskiR05, PodelskiR04LICS, CookPR05} are one of the key approaches proposed to show termination. They both express relationships over program states modeling the progress of variables. From a more pragmatic perspective, showing termination of loops via simple arguments (analysis) has also been studied~\cite{FratantonioMBKV15}. Integrating loop analysis with our approach would help us mitigate precision loss.

\paragraph{\bf Bounded Model Checking} 
Bounded model checking (BMC) is a technique that deals with loops in a systematic manner by simply unrolling (simulating) them~\cite{ClarkeKL04, FalkeMS13, Cordeiro10}. The unrolling process may eventually result in a loop-free code fragment that exactly models the original loop's effect on program variables. Unfortunately, such an approach does not work for loops that are not explicitly bound as the unfolding process will not terminate. Nonetheless, we can combine BMC with our approach to improve our reasoning precision by restricting its application to loops with explicit bounds and apply other techniques to those that are not.

\subsection{Slicing}
\emph{Program slicing} is a technique proposed by Weiser~\cite{Weiser81} to extract a set of statements, called a \emph{slice}, that influence a specified computation of interest, referred to as the \emph{slicing criterion}.  The semantics of the original program are preserved by the slice with respect to the slicing criterion. There has been a tremendous amount of work on slicing and its applications \cite{BinkleyH04}. While the original proposal statically defined a slice, a dynamic variant has been proposed as well~\cite{AgrawalH90}. In the latter, a slice is a set of statements that affect the slicing criterion with respect to a particular input. 

Slicing has been applied to various problems: program debugging~\cite{AgrawalDS93}, testing~\cite{HarmanHLMW07} comprehension~\cite{KorelR98}, re-use~\cite{CanforaLM98}, and re-engineering~\cite{RepsR95}. While the original proposal is syntax-preserving (i.e., the statements of the slice are all taken from the original code), some variants amorphous~\cite{HarmanD97}, allowing changes to the program syntax as long as the program semantics are preserved with respect to the criterion. In the context of software model checking, path slicing was proposed to find statements in a given path that are relevant to show its (in)feasibility~\cite{JhalaM05}. Slicing has also been used to reduce the number of interlivings in event-oriented applications~\cite{BlackshearCS15}, and recently it has been combined with runtime analysis to extract values of variables that make an application difficult to statically analyse~\cite{SiegSME16}.

Our approach shares with slicing the characteristic of relying on dependency analysis. Moreover, our analysis naturally yields sub-Turing slices (i.e., portions of the program that are sub-Turing). We obtain them by simply backtracking paths in the control flow graph of a given method and selecting statements that are not affected by divergent values.

\subsection{Strictness Analysis}
Similar to our approach, strictness analysis has been proposed to track divergence resulting from non-termination and error causing program crashes, such as division by zero. A function is said to be strict if it diverges whenever one of its parameters diverges. A variant of strictness analysis, joint-strictness, takes into account parameter combinations. A function is jointly-strict in a subset of its arguments if it divergences when all the arguments of the subset diverge. Mycroft proposed an approach to approximate the divergence relationship induced by a given function over its parameters and the result it returns~\cite{Mycroft80}. The approach relies on an underlying forward abstract interpretation~\cite{CousotC77}. A backward analysis has been implemented into the Glasgow Haskell Compiler to perform strictness analysis in a demand-driven fashion~\cite{tpda-haskell}. Other forms of strictness analysis have been proposed in the literature. For example, Wadler and Hughes describe several projection-based strictness~\cite{WadlerH87}, such as head-strictness and tail-strictness, refining the original basic definition.    

However, a function being sub-Turing neither entails strictness nor the other way around. Indeed, if a function always diverges regardless of its parameters, it is strict but not sub-Turing. On the other hand, the function \texttt{f(x,y)\{if x return 1 else return y\}} is sub-Turing but not strict. It is sub-Turing as it does not contain any divergent construct. However, it is not strict because in case \texttt{x} is \texttt{true} the function does not diverge even if \texttt{y} diverges.   

\subsection{Pointer Analysis}
\label{sec:pointer_analysis}
Pointer analysis aims at determining the set of memory locations a pointer may refer to during program execution. Two popular pointer analysis that constitute the basis of many other approaches are Steensgaard's \cite{Steensgaard96} Andersen's \cite{Andersen94programanalysis}. While Steensgaard's analysis does not take into account the direction of flow of values induced by assignments, Andersen's approach models assignment direction. Therefore, Steensgaard's technique offers more scalability while Andersen's provides more precision. Das proposed an algorithm lying between Andersen's and Steensgaard's approaches \cite{Das00}. It is scalable and, at the same time, its precision is very close to Andersen's

Lhot{\'{a}}k and Hendren introduced the SPARK framework \cite{LhotakH03} that offers building blocks for implementing various pointer analysis for Java. 

Sridharan \etal proposed a pointer analysis variant which is suitable for environments
with small time and memory budgets \cite{SridharanGSB05}. Their approach is demand-driven, i.e., performs
only the work necessary to answer a query issued by a client.

Instead of applying a pointer analysis, we soundly handle aliases using the variable representative idea inspired by Sundaresan et al~\cite{SundaresanHRVLGG00} (\ref{sec:aliases}). We plan to empirically study the impact of pointer analysis on \ourAnalysis.


\section{Conclusion}

In this paper, we addressed the empirical question of how often a program analysis question has, in practice, an exact solution. To this end, we introduced sub-Turing islands, which are portions of code in which any question of interest is decidable. We provided a formal definition of sub-Turing islands and presented an algorithm for identifying such islands in applications. We have implemented our approach in a tool called \ourTool and applied it to a representative corpus of 1100 Android applications. 

Our empirical study revealed that sub-Turing islands make up 55\% of the methods in the 1100 Android apps studied. These results are not merely of theoretical interest, but have practical ramifications in software engineering. Our findings suggest that we can provide more precise assessments of test coverage; that we can expect more precise assessments of change impact analysis; that we can hope for more precise slices, and thereby, more precise re-use, better comprehension, and better re-engineering interventions.
For example, in the code on which we report, 37\% of runtime-exception guards reside within sub-Turing islands. This means that an exact answer regarding the validity of these guards can be statically determined. 

\label{sec:conclusion}

\bibliographystyle{abbrv}
\bibliography{biblio}



\end{document}